\documentstyle[graphicx,float]{article}

\begin{document}
\title{Are vacuum fluctuations relevant in absorption dynamics?}
\date{}
\author{Pedro Sancho \\ Centro de L\'aseres Pulsados CLPU \\ Parque Cient\'{\i}fico, 37085 Villamayor, Salamanca, Spain}

\maketitle
\begin{abstract}
Vacuum fluctuations play a central role in spontaneous emission. Recently, it has been suggested that these fluctuations could also be fundamental in the absorption dynamics, breaking the superposition inherent to the linear quantum evolution. We analyze the consistency of that proposal with previous results in double spontaneous emission. Moreover, for the case of single absorption by two atoms, we present a test based on the time dependence of the subsequent spontaneous emission patterns, which can  experimentally settle the question. This test is more viable than the original proposal, built on the Casimir effect. Our approach also allows for the comparison between the time scales of vacuum fluctuations as a disentangling mechanism and an emission trigger.
\end{abstract}

Keywords: Vacuum fluctuations; Light-matter interaction; Disentanglement

\section{Introduction}

Vacuum fluctuations are the physical mechanism triggering the process of spontaneous emission \cite{scu}. In the absence of these fluctuations, an excited atom in free space would remain so forever, and the radiation process would never start.

Recently, in \cite{sa}, the possibility that these fluctuations could also play an important role in the process of light absorption has been considered. That author suggests that in an arrangement where two atoms can equally absorb a single photon, the fluctuations could break the two-particle superposition associated with the linear quantum evolution. Instead of the standard assumption of a superposition after the absorption, composed of two two-atom states (each one with one of the atoms in an excited state and the other in the ground one), we would have a mixed state, being the components of the mixture the same of the superposition. This is a different absorption dynamics, which could in principle be experimentally distinguished from the standard one via the Casimir effect. The proposal is interesting because it could illuminate some aspects of the light-matter interaction such as the inherent randomness and, mainly, the role of superposition in the problem.

We analyze in this paper the above proposal from the perspective of previous results in spontaneous emission by pairs of entangled excited atoms. Experimental studies on this process \cite{jap,bel} can be theoretically explained by assuming that the two excited atoms persist in a two-particle superposition state before the spontaneous emission \cite{com,ypr}. The disentanglement does not take place before the first spontaneous emission.

The above argument can be counted as evidence against the \cite{sa} proposal. Nevertheless, more interesting to conclusively resolve the question would be a specific test for the particular setup in \cite{sa}. Such an experimentally feasible test can be formulated by studying the temporal dependence of the patterns of the subsequent spontaneous emission process. They are different depending on whether the initial state is entangled or not. This test of the persistence of two-particle superposition during all the absorption process, and consequently of the potential of vacuum fluctuations to break it, seems to be much more viable than that proposed in \cite{sa}, which is based on the Casimir effect. 

In addition to trigger emission, vacuum fluctuations act as an external environment leading to disentanglement \cite{dod,dis} during the light-matter interaction. Then, our approach also provides a well-suited framework to study this dual role of vacuum fluctuations, in particular the involved time scales. 

\section{Superposition persistence in double emission}

We briefly discuss the arrangement in \cite{sa}. A photon interacts with a pair of identical atoms, and can be absorbed by any of them. According to the linear evolution of quantum theory the process results in a two-atom superposition. Each two-atom state in the superposition is composed of an atom in an excited state, staying the other atom in the ground state. As suggested in that reference, in a way that departs from the standard line of thought, if the vacuum fluctuations were relevant during the absorption process the two-particle superposition could be broken. An equivalent formulation of the idea is that the two-atom entangled state generated by the absorption, disentangles because of the fluctuations before the spontaneous emission.  

The presence of entanglement in excited states and the role it plays in the subsequent spontaneous emission has been studied from both, the experimental and theoretical points of view. The emission patterns of pairs of excited atoms, produced in the photo-dissociation of Hydrogen molecules, were measured in \cite{jap}. The coincidence time spectra, when fitted to an exponential distribution, led to an apparent decay time coefficient different from that of pairs of excited atoms in product states. 
The apparent decay time, and consequently the form of the emission pattern, were dependent on the entanglement present in the system. In order to derive this conclusion, the authors assumed statistical independence between the two atoms during the emission. Clearly, this assumption is doubtful in the context of entangled systems. An alternative method, taking into account the temporal ordering of the two emissions, can overcome that difficulty \cite{com}. Introducing the first and second emission rates one can derive the time-dependent first and second emission patterns. The second one is no loger an exponential distribution, but a combination of two of them. This approach takes into account the statistical correlations between the atoms present in entangled systems, which were ignored in \cite{jap}. A second experiment along the lines in \cite{com} was conducted in \cite{bel}. The main result was that the first and second emission rates are two times and equal to the single-atom emission rate. The values of the emission rates in the new experiment can be mathematically obtained invoking three physical processes: entanglement of the excited atomic states, exchange effects between the atoms (they are identical Hydrogen atoms) and disentanglement of the atoms before the second spontaneous emission \cite{ypr}. 

The important point for our discussion is that the two-particle superposition persists, at least, up to the first emission. This result does not support the view in \cite{sa} of a disentanglement process previous to the emission. However, this argument is not conclusive by two reasons. On the one hand, the photo-dissociation process generating the excited state is not a pure absorption dynamics. The light interacting with the molecule drives two different processes, the dissociation of the molecule and the excitation of the two atoms. On the other hand, the entangled states of the two atoms are not equal in both cases. In \cite{sa} we have the superposition of two two-particle states, each one with only one excited atom. Instead in \cite{jap,bel} both atoms are excited. In spite of these remarks, the result in \cite{ypr} is yet relevant for the proposal in \cite{sa} because the vacuum fluctuations do not break the two-particle superposition of excited states in a closely related context.  

\section{The test}

Although the arguments in the previous section favour the view of a persisting two-particle superposition during the full absorption process it would be desirable to have a specific viable test for the case of a single absorption. We propose in this section such a test for identical atoms.

Next, we describe the setup (see Fig.1). Two identical atoms (later we justify the need of using indistinguishable particles) interact with a light beam, which is weak enough in order to only one absorption be possible in each repetition of the experiment. The atoms after the absorption, depending on using the standard approach or the alternative \cite{sa} one, can be in  two-boson/fermion superposition pure states or in two-boson/fermion mixed ones. We shall evaluate and compare the spontaneous emission patterns of both approaches in order to physically distinguish them.

The initial state of the atoms is
\begin{equation}
|\Psi _0>=N_0 (|\psi_0>_1|\phi_0>_2 \pm |\phi_0>_1|\psi_0>_2 )|g>_1|g>_2
\end{equation} 
with $\psi_0$ and $\phi_0$ the two initial one-particle center of mass (CM) states and $g$ denoting the electronic ground one. The subscripts $1$ and $2$ are the labels of the two atoms. In the double sign $\pm$ the upper one holds for bosons and the lower one for fermions. The normalization coefficient reads $N_0=(2 \pm 2|<\psi_0|\phi_0>|^2)^{-1/2}$. The scalar product $<\psi_0|\phi_0>$ measures the overlap between the two atoms. As we shall discuss afterwards, in order the test to work, that overlapping cannot be negligible.

\begin{figure}[H]
\center
\includegraphics[width=8cm,height=7cm]{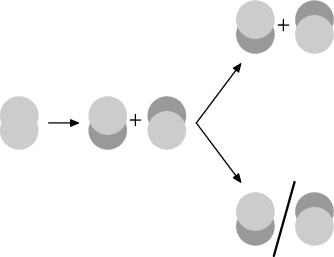}
\caption{Sketch of the scheme. The initial preparation consists of two overlapping identical atoms. After the absorption of a photon it evolves to a superposition (+) of two two-atom states, representing the darker grey the one-particle state of the absorbing atom. Later, previous to the spontaneous emission, we consider two different evolutions. The upper one, corresponds to the standard linear approach, where the superposition persists. The lower one, refers to the proposal in \cite{sa}, in which the superposition breaks because of the vacuum fluctuations. In this case we have a mixture (/) of the same states of the superposition. Comparing the spontaneous emission rates of both evolutions we can experimentally test them.}
\end{figure}

The atoms interact with the one-photon beam of light, represented by the state $|1>_{EM}$, with $EM$ referring to the electromagnetic field of the light. Then the full initial state $|\Psi _0>|1>_{EM}$ can evolve in three different ways, no absorption, absorption by the atom in state $\psi_0$, and absorption by the atom in state $\phi_0$. The two last alternatives lead to the states 
\begin{equation}
|\psi^*_{abs}> = \frac{1}{\sqrt 2}( |\psi^*>_1|e>_1|\phi>_2|g>_2 \pm |\phi>_1|g>_1  |\psi^*>_2|e>_2  )
\end{equation}
and 
\begin{equation}
|\phi^*_{abs}> = \frac{1}{\sqrt 2}( |\psi>_1|g>_1|\phi^*>_2|e>_2 \pm |\phi^*>_1|e>_1  |\psi)>_2|g>_2  )
\end{equation}
The CM one-particle states after the absorption are $\psi^*$ and $\phi^*$, denoting the superscript $*$ the effect of the recoil. The states $\psi$ and $\phi$ are the free evolving spatial states of the atoms that do not absorb light. They are the states that are obtained using the interaction picture. On the other hand, $e$ refers to the excited electronic state. As the two absorption alternatives are indistinguishable we must add both amplitudes, or equivalently, we have a superposition of both states:
\begin{equation} 
|\Psi_{abs}>=N_{abs}(|\psi^*_{abs}>+|\phi^*_{abs}>)
\end{equation}
with the normalization coefficient 
\begin{equation}
N_{abs}=(2 \pm 2 Re(<\psi^*|\phi^*><\phi|\psi>))^{-1/2}
\end{equation}
In order to concentrate on the states with absorption, we postselect them and discard the cases without absorption. 

The presence of a multi-particle superposition is the standard approach to absorption. If, instead, we adopt the alternative view of \cite{sa} where the superposition is broken by the vacuum fluctuations, at the end of the absorption process we would have a mixture with equal weights of states $\psi ^*_{abs}$ and $\phi ^*_{abs}$.

When the absorption process is completed, the spontaneous emission one begins. There are two alternatives for the process, associated with the emission by an atom in state $\psi^*$ or $\phi^*$, and represented by the states    
\begin{equation}
|\psi_{\Omega}>=N_{\psi_{\Omega}}(|\psi_{\Omega}^{sp}>_1|\bar{\phi}>_2 \pm |\bar{\phi}>_1|\psi_{\Omega}^{sp}>_2)|g>_1|g>_2
\end{equation}
and
\begin{equation}
|\phi_{\Omega}>=N_{\phi_{\Omega}}(|\bar{\psi}>_1|\phi_{\Omega}^{sp}>_2 \pm |\phi_{\Omega}^{sp}>_1|\bar{\psi}>_2)|g>_1|g>_2
\end{equation}
with the normalization coefficients $N_{\psi_{\Omega}}=(2\pm 2|<\psi_{\Omega}^{sp}|\bar{\phi}>|^2)^{-1/2}$ and $N_{\phi_{\Omega}}=(2\pm 2|<\phi_{\Omega}^{sp}|\bar{\psi}>|^2)^{-1/2}$. The superscript $sp$ indicates the CM state after the recoil associated with the spontaneous emission.  The spatial states $\bar{\psi}$ and $\bar{\phi}$ are the free evolution (corresponding to the interaction representation) of $\psi$ and $\phi$. The parameter $\Omega$ refers to the direction of emission of the photon. The spontaneous emission is a random process that can take place in any direction. We only consider the case where we post-select the cases with that fixed direction. 

After the absorption of the photon the influence of the vacuum fluctuations in the problem begins. The standard view is that the fluctuations trigger the spontaneous emission but without affecting to the superposition, that is, preserving the pure character of the state $\Psi_{abs}$. In contrast, in \cite{sa} the fluctuations break the superposition before the emission, leading to a mixture of $\psi_{abs}^*$ and  $\phi_{abs}^*$. We separately evaluate these two paths to emission, showing that depending on the role associated to vacuum fluctuations in the problem, we can have different emission patterns.

We analyze first the standard approach, where the superposition persists. The two emission alternatives are indistinguishable: the frequency and direction of the emitted photon is the same in both cases. The assumption of a large overlap between the two atoms, which persists after the small recoil associated with the spontaneous emission, precludes the possibility of distinguishing which of the two atoms emitted the photon. The fact that the alternatives cannot be distinguished and, consequently, the probability amplitudes must be added, is equivalent to the persistence of the two-particle superposition, which is represented by the pure state  
\begin{equation}
|\Psi_{\Omega}^{sp}> =N_{\Omega}^{sp}(|\psi_{\Omega}>+|\phi_{\Omega}>)
\end{equation}
with
\begin{eqnarray}
N_{\Omega}^{sp}=(2+2Re(<\psi_{\Omega}|\phi_{\Omega}>))^{-1/2} = \nonumber \\ 
(2+4N_{\psi_{\Omega}}N_{\phi_{\Omega}}Re(<\psi_{\Omega}^{sp}|\bar{\psi}><\bar{\phi}|\phi_{\Omega}^{sp}> \pm <\psi_{\Omega}^{sp}|\phi_{\Omega}^{sp}><\bar{\phi}|\bar{\psi}>))^{-1/2}
\end{eqnarray}

Then the probability amplitude for spontaneous emission is
\begin{equation}
{\cal M}_{\Omega}(t)=_{EM}<1|<\Psi_{\Omega}^{sp}|\hat{U}(t)|\Psi_{abs}>|0>_{EM}
\end{equation}
with $\hat{U}$ the full, light field plus atoms, evolution operator. We take $t=0$ as the time at which the emission process begins.

As there is no interaction between the atoms, the evolution operator can be factored as $\hat{U}=\hat{U}_1 \otimes \hat{U}_2$, with $\hat{U}_j, j=1,2$, describing the interaction of each atom with the field. At first order of perturbation theory the operator can be expressed as $\hat{U}_j \approx \hat{I}_j-it\hat{H}_j/\hbar$, with $\hat{I}_j$ the identity operator and $\hat{H}_j$ the usual electric-dipole one-atom interaction Hamiltonian in the interaction representation \cite{Lou}.

The explicit form of the matrix element is
\begin{eqnarray}
{\cal M}_{\Omega} =-\frac{itD}{\hbar} \frac{N_{abs}N_{\Omega}^{sp}}{\sqrt 2}[N_{\psi_{\Omega}}(2+2<\psi_{\Omega}^{sp}|\bar{\psi}><\bar{\phi}|\phi_{\Omega}^{sp}> \pm \nonumber \\
2|<\psi_{\Omega}^{sp}|\bar{\phi}>|^2 \pm 2<\psi_{\Omega}^{sp}|\phi_{\Omega}^{sp}><\bar{\phi}|\bar{\psi}>)   +  N_{\phi_{\Omega}} (2+ \nonumber \\
2<\bar{\psi}|\psi_{\Omega}^{sp}><\phi_{\Omega}^{sp}|\bar{\phi}> \pm 
2|<\phi_{\Omega}^{sp}|\bar{\psi}>|^2 \pm 2<\phi_{\Omega}^{sp}|\psi_{\Omega}^{sp}><\bar{\psi}|\bar{\phi}>)  ]
\end{eqnarray}
with $D=D_0{\bf e}. <g|\hat{\bf D}|e>$ the scalar product of the unit polarization vector of the light and the expectation value of the electric-dipole moment operator of the atom between states $e$ and $g$ . $D_0$ contains all the constant coefficients present in the interaction Hamiltonian.

The absorption rate $\Gamma _{\Omega}$ can be obtained in the standard way via the Fermi golden rule \cite{Lou}. It is proportional to $|_{EM}<1|<\Psi_{\Omega}^{sp}|\hat{H}(t)|\Psi_{abs}>|0>_{EM}|^2$ and, consequently, can be easily derived from  ${\cal M}_{\Omega}$. The constant coefficients arising from the application of the rule are absorbed in $D_0$. Assuming that initially there are $n_0^{\Omega}$ excited pairs of atoms whose emission takes place in the direction $\Omega$ (remember that only this case is postselected, the rest of excited atoms emitting in other directions are not taken into account), the usual decaying law gives the number of atoms that remain excited at a given time $n_{exc}(t)=n_0^{\Omega} \exp(-\Gamma_{\Omega} t)$. Equivalently, the number of emitted photons in the direction $\Omega$ at a given time is
\begin{equation}
n_{emi}^{\Omega}(t)=n_0^{\Omega}(1-\exp(-\Gamma _{\Omega}t))
\end{equation} 
Next, we consider the non-standard approach where we have a mixture with equal weights of $\psi _{abs}^*$ and $\phi _{abs}^*$. Now, there are not two alternatives in each component of the mixture. In one of the components the only alternative is $\psi _{abs}^* \rightarrow \psi_{\Omega}$, and in the other $\phi _{abs}^* \rightarrow \phi_{\Omega}$. The matrix elements associated with the two cases are 
\begin{eqnarray}
{\cal M}_{\Omega}^{\psi}(t)=_{EM}<1|<\psi_{\Omega}|\hat{U}(t)|\psi_{abs}^*>|0>_{EM} = \nonumber \\
- \frac{itD}{\hbar}\frac{N_{\psi}}{\sqrt 2}(2 \pm 2|<\psi _{\Omega}^{sp}|\bar{\phi}>|^2)
\end{eqnarray} 
and 
\begin{eqnarray}
{\cal M}_{\Omega}^{\phi}(t)=_{EM}<1|<\phi_{\Omega}|\hat{U}(t)|\phi_{abs}^*>|0>_{EM} = \nonumber \\
- \frac{itD}{\hbar}\frac{N_{\phi}}{\sqrt 2}(2 \pm 2|<\bar{\psi} |\phi_{\Omega}^{sp}>|^2)
\end{eqnarray} 
As for the standard case, using these probability amplitudes we can obtain the absorption rates for the two alternatives, 
$\Gamma_{\Omega}^{\psi}$ and $\Gamma_{\Omega}^{\phi}$. As the two terms of the mixture have the same weight we easily arrive to the number of photons emitted in the direction $\Omega$ by the mixture
\begin{equation}
n_{emi}^{\Omega , mix}(t)=n_0^{\Omega}(1-\frac{1}{2}\exp(-\Gamma _{\Omega}^{\psi}t) -\frac{1}{2}\exp(-\Gamma _{\Omega}^{\phi}t))
\end{equation}
The two emission distributions are different. We explicitly represent in Fig. 2 the two cases in order to see the differences in a graphical way. The absorption rate can be expressed as $\Gamma _{\Omega}=\Gamma _0 f$ where $f$ contains the normalization coefficients and scalar products, with $\Gamma_0=|D|^2/\hbar^2$. Similar expressions hold for $\Gamma _{\Omega}^{\psi}$ and $\Gamma _{\Omega}^{\phi}$. We take $\Gamma_0=1$ and use the scalar products $<\psi|\phi>=0.7$ (the free parameter of the representation), $<\psi^*|\phi^*>=(0.9+0.1<\psi|\phi>)<\psi|\phi>$, $<\psi_{\Omega}^{sp}|\phi_{\Omega}^{sp}>=0.9<\psi^*|\phi^*>$, $<\psi_{\Omega}^{sp}|\psi>=0.9=<\phi_{\Omega}^{sp}|\phi>$, $<\psi_{\Omega}^{sp}|\phi>=(0.8+0.1<\psi|\phi>)<\psi|\phi>$ and $<\phi_{\Omega}^{sp}|\psi>=(0.8+0.1<\psi|\phi>)<\psi|\phi>$. With this choice we guarantee that all the recoils are small and that in the limit $\psi = \phi$ we have $\psi^*=\phi^*$,....
\begin{figure}[H]
\center
\includegraphics[width=10cm,height=7cm]{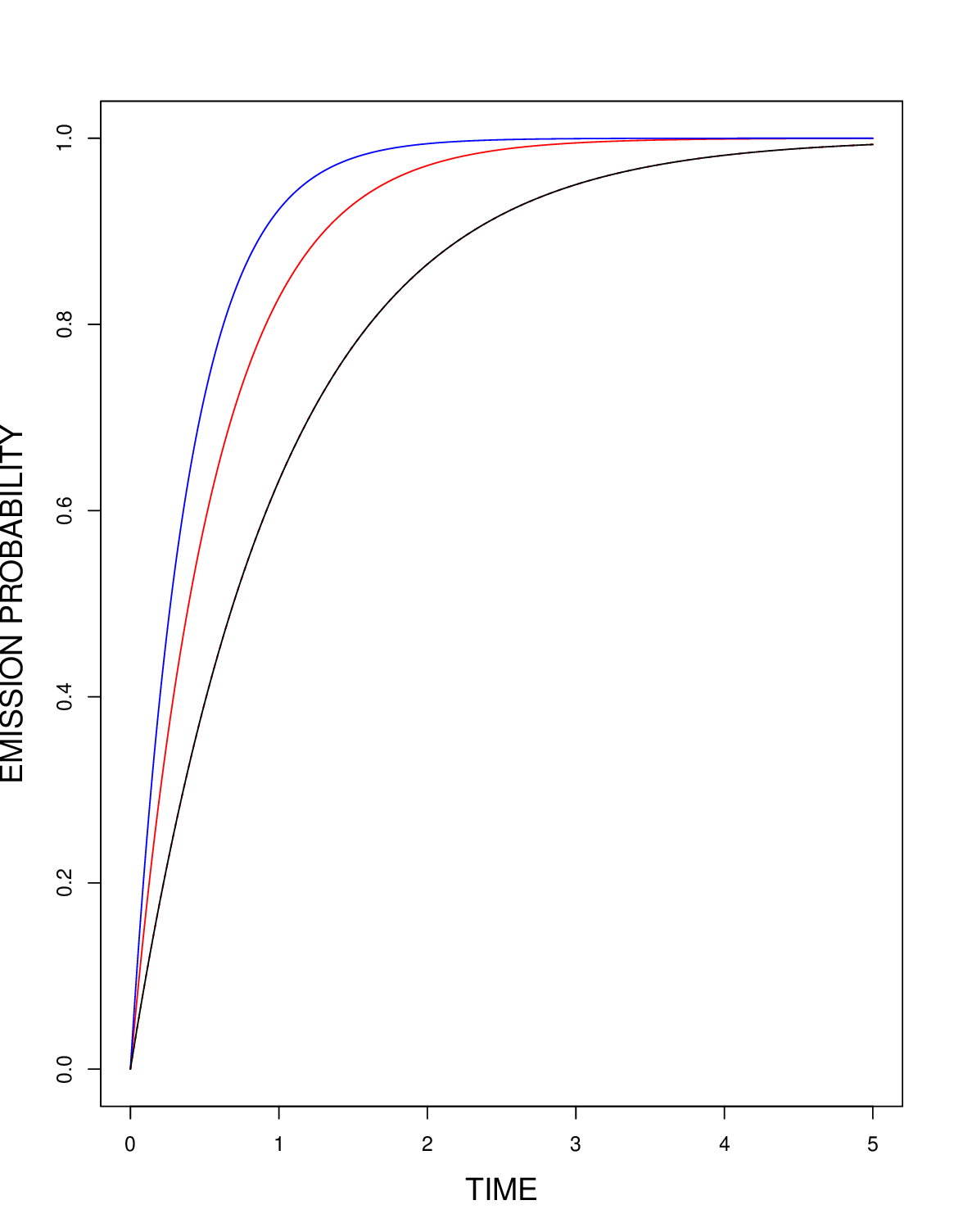}
\caption{Emission probability $n_{emi}^{\Omega}/n_0^{\Omega}$ versus time $t$ in units of $\Gamma_0^{-1}$. The red, blue and black lines correspond respectively to initial excited states of a boson superposition, fermion superposition and (boson or fermion) mixture. The mixture corresponds to the case of vacuum fluctuations breaking the superposition generated by the absorption process.}
\end{figure}

The fermions in two-particle superposition emit faster than bosons, and both than mixtures. As can be easily corroborated analytically, there is no difference between mixtures of fermions and bosons. In principle, these differences in the emission patterns are experimentally measurable, making possible to test if the vacuum fluctuations can break the superposition caused by the photon absorption. 

\section{Discussion}

We have critically analyzed the proposal in \cite{sa} for an alternative absorption dynamics. Previous results in excited two-atom
states with double spontaneous emission suggest, contrarily to \cite{sa}, the persistence of the two-particle superposition during the full absorption process. The main result of the paper is the proposal of a realistic test to experimentally settle the question.

The test suggested here seems to be more viable than the setup based on the Casimir effect described in \cite{sa}. In effect, in the last case the atoms must be placed between parallel conductive plates whose separation can be controlled. For different separations we have different intensities of the vacuum fluctuations, leading to different values of the persistence of the atomic superposition. When the emission occurs during the persistence of the superposition state we have a quantum interference pattern, associated with the existence of two indistinguishable alternatives in the superposition. The interference dissapears for emissions taking place after the superposition breaking. If the duration of the interval showing interference patterns varies with the plates separation we will have an unequivocal proof of the presence of the breaking mechanism. Although in \cite{sa} there is not an estimation of the duration of the interference effects as a function of the plates separation, it is clear that this is a very tiny effect. The experiment is very demanding. In contrast, in our proposal it is only necessary to measure the temporal dependence of the emissions, by far a much simpler task.     

We emphasize that our test is valid for identical particles with a non-negligible overlap. In the case of distinguishable atoms the spontaneous emission would take place at different frequencies and, consequently the emission probability amplitudes would do not add but the probabilities. It is simple to verify that in this case the emission patterns are equal for excited atoms in two-particle superposition and for mixtures. Only in the case that the emission frequencies of two distinguishable atoms after including the radiative broadening were indistinguishable, a similar test for non-identical atoms would be viable. Moreover, the overlap between identical atoms must be different from zero. If the overlap would be negligible the atoms could be treated as distinguishable ones (using the CM position as a physical label). It would be possible to associate the emission to one of the two positions, making the two processes distinguishable. Again there would not be difference between the emission patterns of excited atoms in two-particle superposition and mixtures. 

The test has been derived for a particular direction of emission. When we consider all the directions, as they are distinguishable, we must add probabilities not probability amplitudes. The absorption rate is the continuous sum of all the rates associated with these probabilities. We do not consider here this problem because the integration of all the rates, with different values of the scalar products depending on the emission direction, seems to be very complex.  

Another topic where vacuum fluctuations are relevant is that of correlations outside the light cone in quantum electrodynamics. Because of the fluctuations, two systems placed at causally disconnected regions can become statistically correlated \cite{bis,val}, as experimentally shown in \cite{nc}. As a natural extension of these ideas, the possibility of generating entanglement from the vacuum has been also considered \cite{rez}. It seems plausible to extend the ideas presented in this paper to design tests of the vacuum correlations using excited atoms located at causally disconnected points.

As signalled in the introduction, vacuum fluctuations play a dual role during the light-matter interaction. They trigger the spontaneous emission, but at the same time they drive a disentanglement process. The last role has been extensively studied in the literature. In disentanglement theory, an external environment interacting with an entangled system acts as an entanglement breaking mechanism \cite{dod,dis}. In particular, in the case of spontaneous emission, where the vacuum fluctuations are the only environment present, there are detailed results on the time dependence of the process, sometimes occurring in a finite time instead of an asymptotic way \cite{ebe}. In the language of disentanglement theory, the proposal in \cite{sa} is equivalent to assume that the time scale of vacuum fluctuations as a disentangling mechanism is shorter than that as an emission trigger. The test here proposed can be seen as a way to compare both time scales.

\end{document}